\documentclass[referee]{jfm}
\usepackage{graphicx}

  \title[Impact of floats on water]{Impact of floats on water}
  \author[N. de Divitiis and  and L. M. de Socio]
          {N.\ns  d\ls e\ns  D\ls I\ls V\ls I\ls T\ls I\ls I\ls S\ls 
          \and L.\ns M.\ns  d\ls e\ns S\ls O\ls C\ls I\ls O\ls }
  \affiliation{University of Rome ``La Sapienza'', 00184 Rome, Italy }
  \date{}

\begin{document}           
\baselineskip=0.950cm

\maketitle              

\newcommand{\no}{\noindent}
\newcommand{\be}{\begin{equation}}
\newcommand{\ee}{\end{equation}}
\newcommand{\bea}{\begin{eqnarray}}
\newcommand{\eea}{\end{eqnarray}}
\newcommand{\bc}{\begin{center}}
\newcommand{\ec}{\end{center}}

\newcommand{\calr}{{\cal R}}
\newcommand{\calv}{{\cal V}}

\newcommand{\bff}{\mbox{\boldmath $f$}}
\newcommand{\bfg}{\mbox{\boldmath $g$}}
\newcommand{\bfh}{\mbox{\boldmath $h$}}
\newcommand{\bfi}{\mbox{\boldmath $i$}}
\newcommand{\bfm}{\mbox{\boldmath $m$}}
\newcommand{\bfp}{\mbox{\boldmath $p$}}
\newcommand{\bfr}{\mbox{\boldmath $r$}}
\newcommand{\bfu}{\mbox{\boldmath $u$}}
\newcommand{\bfv}{\mbox{\boldmath $v$}}
\newcommand{\bfx}{\mbox{\boldmath $x$}}
\newcommand{\bfy}{\mbox{\boldmath $y$}}
\newcommand{\bfw}{\mbox{\boldmath $w$}}
\newcommand{\bfk}{\mbox{\boldmath $k$}}

\newcommand{\bfA}{\mbox{\boldmath $A$}}
\newcommand{\bfD}{\mbox{\boldmath $D$}}
\newcommand{\bfI}{\mbox{\boldmath $I$}}
\newcommand{\bfL}{\mbox{\boldmath $L$}}
\newcommand{\bfM}{\mbox{\boldmath $M$}}
\newcommand{\bfS}{\mbox{\boldmath $S$}}
\newcommand{\bfT}{\mbox{\boldmath $T$}}
\newcommand{\bfU}{\mbox{\boldmath $U$}}
\newcommand{\bfX}{\mbox{\boldmath $X$}}
\newcommand{\bfY}{\mbox{\boldmath $Y$}}
\newcommand{\bfK}{\mbox{\boldmath $K$}}

\newcommand{\bfrho}{\mbox{\boldmath $\rho$}}
\newcommand{\bfomega}{\mbox{\boldmath $\omega$}}
\newcommand{\bfeps}{\mbox{\boldmath $\varepsilon$}}
\newcommand{\itPsi}{\mbox{\it $\Psi$}}
\newcommand{\itPhi}{\mbox{\it $\Phi$}}
\newcommand{\bint}{\mbox{ \int{a}{b}} }
\newcommand{\ds}{\displaystyle}        

                

\begin{abstract}

\no The impact of a wedge-shaped body on the free surface of a weightless inviscid incompressible liquid is considered. 
Both symmetrical and unsymmetrical entries at constant velocity are dealt with. The differential problem corresponds 
to the physico-mathematical model of a distribution of potential singularities and, in particular, the flow singularities 
at the ends of the wetted regions are represented by sinks. A conformal transformation of the flow field is adopted and 
the unknown intensities of the discontinuities are found by an optimization procedure, together with the solution of the 
nonlinear free-surface problem. The flow separation at a sideslip is also considered.

\end{abstract}

\bigskip

\section{Introduction}

Recently a renewed attention has been given to the  hydrodynamic action on floats during their entry into water.
Apart from the challenging mathematical aspects of the problem,  this is due to the renewed  interest in large seaplanes
in the aircraft world and in very fast marine vehicles in ship building.  One of the first models for dealing with the
\begin{figure}
\vspace{0.mm}
\centering
\includegraphics[width=0.90\textwidth]{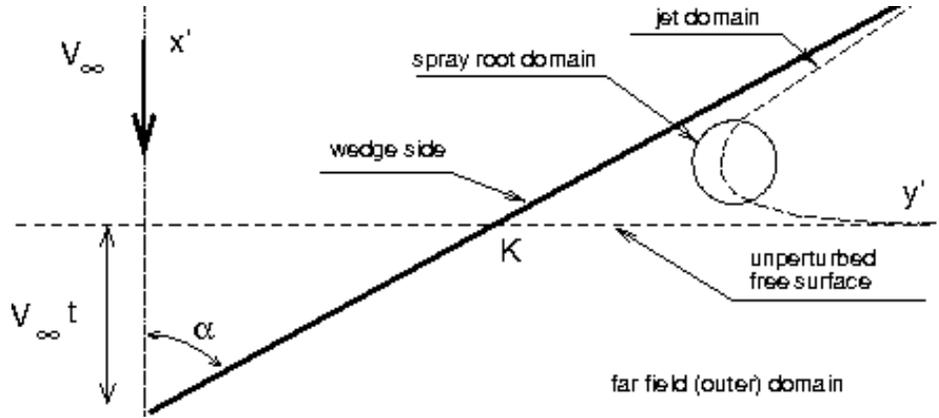}
\caption{Proposed subdomains.}
\end{figure}

hydrodynamics of a seaplane
 just after its impact on the water surface was proposed by  \cite{Karman29}.
Much later, significant contributions to the solution of the  slamming problem came from naval architects
when the full picture of a ship slamming into water began to be considered in all its aspects from the
hydrodynamical phenomena to the structural aspects. In this framework the mechanism of the solid surface interaction
with the liquid was experimentally observed and theoretically modelled, including such effects as the air cushion formation,
vortex generation and hydroelasticity.

\cite{Faltinsen93} presents an good review of the foundations of the slamming problem and their connections
with the seakeeping of vessels, where the stresses induced by a water impact can play a very important role. On this
last point see, for instance, some recent analytical and numerical solutions in \cite{Iafrati97} and \cite{Carcaterra98}.

Due to the complexity of the impact phenomenology, in most studies it is assumed that the liquid is inviscid 
and there are no effects of gravity. This means that one of the characteristic parameters which govern the physics of 
the problem, namely the Froude number, is supposed to be much greater than unity, $Fr = V_{\infty}/g t \gg 1$, and the Reynolds
number $Re = {V_{\infty}}^2 t /\nu \rightarrow \infty$, where $V_{\infty}$ is the entry velocity, $g$ acceleration due to the gravity,
 $t$ time and $\nu$ fluid viscosity.  This  provides a limit on the validity of the results
 of the theory in terms of the time $t$, which must be shortly after the impact, and of the impact speed $V_{\infty}$ which must be large. 

In most of articles, such as those by \cite{Karman29}, \cite{Wagner32}, \cite{Dobrovolskaya69}, \cite{Cointe91}, \cite{Zhao93}, 
\cite{Faltinsen97}, the reference shape of the float is assumed  to be a wedge that  is usually supposed to have its symmetry
 plane normal to the free water surface.

 A number of papers deal with the entry of blunt bodies. When this is the case, a simple approach 
is to assume that the first impact is that of a flat plate (Moghishi $\&$ Squire 1981, Cointe 1987, Howison, Ockendon $\&$ Wilson  1991).
 It was shown that Wagner's solution for a wedge-shaped body also applies to arbitrary blunt bodies, provided that 
the wetted length is properly computed. Also interesting is the fact that the asymptotic solution for the wedge entry problem is 
a particular example of the more general asymptotic approach for blunt bodies.
Present trends in naval research follow more complex procedures than the flat plate approach when dealing with
blunt geometries but they are beyond the scope of this paper.

 The symmetric impact of a  wedge-shaped float can be divided in two phases. Initially  the point of contact
$K$ (figure 1) between the wall and the unperturbed free surface may move at a velocity $V_K$ greater than the speed of 
sound in the  liquid $c_w$. In particular $V_K = V_{\infty}/\cos \alpha$, where $\alpha$ is the wedge semi-angle  and $V_{\infty}$ is the 
entry velocity.  After the very first instants, however, $V_K$ slows down to subsonic values.  The supersonic and the subsonic phases 
are, of course, treated in different ways. When a supersonic phase occurs it is generally dealt with within an acoustic approximation 
(see {\cite{Skalak66}}, {\cite{Korobkin92}})  which is sufficiently accurate in  applications where the Mach number of point $K$, 
namely $V_K/c_w$, is  a little greater  than unity.  

The subsonic case of an incompressible fluid, which  is the subject of the present paper, has a series of
interesting features associated with the complicated configuration of the flow field and the nonlinear aspects of the mathematical 
problem due to the presence of a free surface.

 Figure 1 shows the subdomains into  which the water region
 can be divided according to Wagner's (1932) first ideas, the first quantitatively correct 
version of \cite{Howison91} and then \cite{Cointe91}, \cite{Zhao93} and  \cite{Faltinsen97}; this
 figure helps  in understanding the physical characteristics of the field.
 The  sketch also indicates the reasons for some of the  approximate
 solution procedures which were adopted in the past. In particular one can see the presence of  
 a close field with sizeable effects  on the  impact, a far field of negligible perturbations, and
  two lateral jets. These jet subdomains, first described by \cite{Wagner32}, \cite{Howison91} and
 then by \cite{Cointe91} and others, are the regions where portions of the liquid initially run close
 to the wall  and then lose their continuum fluid identities and from mists,  sprays or, 
 in general,  two-phase flows.

When considering the existing literature, recall that solutions to the subsonic incompressible problem have appeared frequently,
using different either approximate or numerical approaches. For a constant entry velocity  into an inviscid and weightless
 liquid, approximate analytical solutions have been proposed which are  similar with respect to the  time $t$. In particular the 
two-dimensional Laplace equation for the velocity  potential was initially solved by \cite{Wagner32} in the case of a wedge of very 
small  vertex angle. Much later \cite{Dobrovolskaya69}  reduced the problem of the complex potential to a nonlinear singular 
integral equation which was solved by a method of successive approximations.  Subsequently, a solution to this equation was found by 
\cite{Zhao93} by a nonlinear  boundary elements method. These authors also pointed some the errors in \cite{Dobrovolskaya69} data.

On the other hand, for large values of $\ds \alpha < {\pi}/{2}$, \cite{Korobkin88}, used a variational 
approach, which was later followed by \cite{Howison91}, for an asymptotic  analysis. 
In addition \cite{Fraenkel97} carried  out an asymptotic analysis for $\ds \alpha = {\pi}/{2}-\epsilon$, after a 
conformal transformation  of the field. Almost at the same time \cite{Fontaine97} published a summary of the approximate 
 (for large $\alpha$) results and proposed composite solutions which are based on a division  of the flow field analogous to the one 
in figure 1.

A conformal mapping method, which involves the Wagner's function, was used by \cite{Hughes72} to solve the water entry problem
 of a wedge by a mixed analytical and numerical procedure. The method reduces the problem to the calculation of a mapping 
function for the hodograph.

The impact may not be symmetrical for different reasons:  the symmetry plane of the wedge
is normal to the free surface whereas  the entry velocity is not; or the entry velocity is normal to the free surface 
and the symmetry plane is not.
A few authors speculated about possible approaches to the situation of a wedge plunging into the water at a sideslip 
angle, but no calculated solutions were  presented   (Wagner 1932; Dobrovolskaya 1969). The particular case of flow separation
 was considered by \cite{Zhao97} in the framework of a simulation study on the entry of two-dimensional bodies of arbitrary 
cross-sections. They dealt with the problem of flow separation from knuckles or fixed separation points on both sides of symmetric
bodies. In particular, the Kutta condition is applied at a separation point. 
In the second case of an asymmetric water entry, where a wedge plunges normal to the free surface but with the velocity vector at 
an angle with respect to the symmetry plane, \cite{Toyama93} presented a solution obtained through  a finite element method.

\cite{Morgan94}, gave an explanation of  Trefethen $\&$ Panton's (1990) observation that an impact splash is largely 
independent of the horizontal speed of the impacting body, when the horizontal velocity component is comparable with
the downward velocity.

Here, after proposing a model of the wedge slamming problem which is based on a suitable distribution of the velocity potential,
we find excellent solutions of that model
for the flow field following an optimization procedure for solving an algebraic set of equations. The shape of the free 
surface and the pressure coefficient distribution along the  wetted walls will be considered in cases of both symmetrical 
and  non-symmetrical  impact. In the latter case, in the presence of sideslip, we obtain the solution in the two situations of a flow which is either
 attached to both faces of the float or  separates downstream from the leading edge. Moreover a condition for flow
separation will be introduced.
 
As in some of the cited references, conformal transformations will be used, although in our case  the external hydrodynamical problem in 
the semi-infinite physical plane   will be transferred to part of the interior of a circle in the transformed plane.  The solutions will be
 compared with the existing ones where available.

\section {\bf  Analysis of the two-dimensional impact}

Let us again consider figure 1. A preliminary observation suggests a reasonable mathematical model  of  the physics of slamming,
which is based on a distribution of singularities in a steady potential field.  For a constant velocity $V_{\infty}$, an appropriate
transformation of the $x'$- and $y'$-coordinates into the dimensionless ones, $x = \ds  {x'}/{V_{\infty} t}$  and 
$y= \ds {y'}/{V_{\infty} t}$, makes the problem similar with respect to the time $t$ and steady in  the new  coordinate 
 plane ($x, y$). It  is convenient to take the body at rest while the water level moves upward, and this assumption justifies why 
the calculated iso-$\psi$ lines, in the present model, end at the free surface. 

At the walls the normal component of the flow velocity is zero whereas the perturbing effects of the body impact  vanish in all 
directions at an infinite distance from the wedge. Let the free surface be represented in the plane $(x, y)$ by the line 
$ x=\bar{x}(s), y=\bar{y}(s)$. The unknown shape of this curve, which makes the differential problem nonlinear, is to be determined by imposing that the  pressure  coefficient 
 $C_p$  at the free surface must be zero and that its perturbation from the unperturbed straight line goes to zero as $s$ goes to
 infinity.  As we will see, both these conditions can be satisfied provided that  a jet of finite flow rate, also to be determined, 
is present close to the float, on each side.
\begin{figure}
\vspace{0.mm}
\centering
\includegraphics[width=0.90\textwidth]{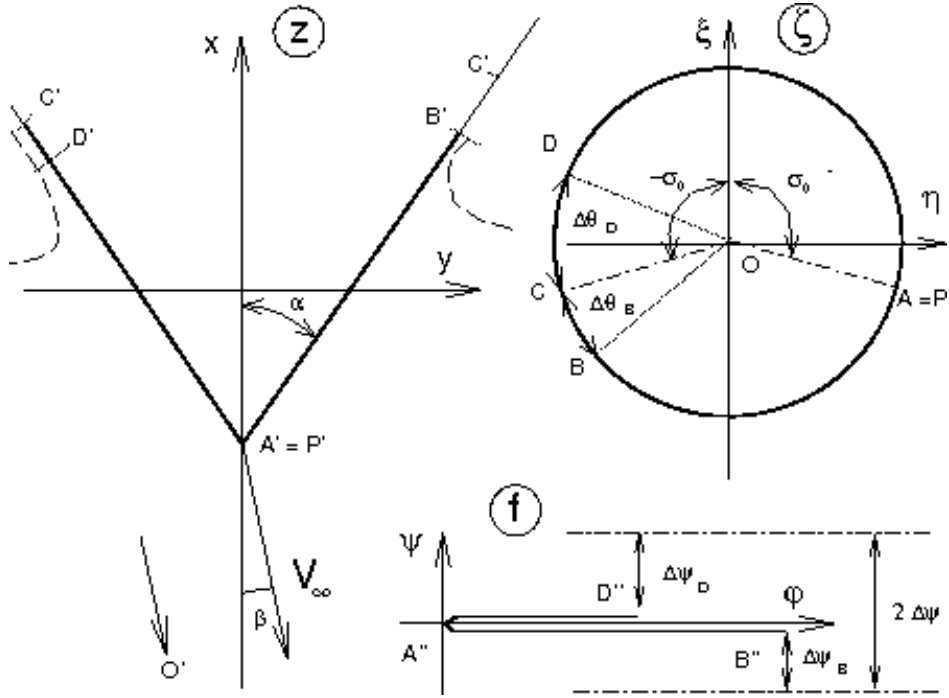}
  \caption{Conformal mapping (attached flow).}
\end{figure}
A contact iso-$\psi$ line separates the jet from the rest of the flow.
For the case where  the flow separates from one side some further considerations will be presented later.

As already observed in previous works, the potential flow field in an incompressible weightless fluid presents similar
solutions in the transformed coordinates $x, y$, provided that the velocity potential $\Phi$ and the streamfunction $\Psi$ are 
expressed in the dimensionless transformed forms  $\varphi$ and $\psi$:
\bea
 \itPhi(x',y',t) = V^2_{\infty} \ t \ \varphi (x, y); \ \ \ \itPsi(x',y',t) = V^2_{\infty} \ t \ \psi (x, y).  
\label{similar}
\eea
Then the velocity $(u', v')$ of the flow field, expressed as the gradient of $\Phi$,  has the form
\bea
(u', v') \equiv (\frac{\partial \itPhi }{\partial x'}, \frac{\partial \itPhi }{\partial y'}) = V_{\infty} {\bf v},
\label{similarV}
\eea 
with $\ds {\bf v} \equiv (u, v)= ({\partial \varphi }/{\partial x}, \ {\partial \varphi }/{\partial y})$.
 The  differential  problem represented by the Laplace equation and associated boundary conditions will be solved by taking 
advantage of the conformal transformations in the situations which are sketched in figures 2 and 3. Note that, for convenience, 
we have assumed a circle in the complex plane  for the transformed physical water region.

Figure 2 shows a flow field where the fluid is attached to both sides of the wedge in an unsymmetrical water entry, of which 
a symmetrical situation is a particular case. Figure 3 corresponds to the circumstance where the flow separates from one 
side when the float enters at a sideslip angle $\beta$.  

Let $z = x +$ i $y$ be the physical plane and $\zeta = \xi +$ i $\eta$  the complex plane onto which $z$ is conformally transformed,
 with the walls of the wedge becoming arcs of a circle the centre of which, $O$, corresponds to the upstream infinity.  Let the complex
 potential $f =\varphi +$ i $\psi$ be defined on  $\zeta$. Corresponding points of the two planes are represented by  the same capital 
letters with a prime  for the points on $z$.  In all cases, points $A'$ and $P'$ refer to the stagnation point and to wedge vertex,
respectively, and so do their transformed representations $A$ and $P$. Between $B$ and $O$ and between $D$ and $O$ lie the unknown
lines which represent the two branches of the free surface.

The general model of the flow field corresponds to the sum of a number of singularities of the velocity potential, the intensities 
of which are to be determined in accordance with the physical conditions.

As already mentioned the conditions on the physical plane $z$ correspond to a  vanishing normal component of the velocity along  the lines $A'B'$ and 
$A'D'$, whereas $C_p=0$ at all the points of the free surface, including $B'$ and $D'$. In particular, this last point represents
an excellent assumption for small and intermediate deadrise angles and is still a good approximation for $\alpha$
as low as $9 ^{\circ}$.

The meanings and the locations of the singularities are quite easily understood.
There is a doublet at $O$ for the translational potential. 
More attention has to be paid to the singularities which provide the simulation of the lateral jets.
In this case each jet is represented by a sink and a distorted doublet at, for example, point $C'$, in such a way that the necessary 
jump of the streamfunction $\Delta \psi$ is obtained, while the condition $C_p =0$ is satisfied at both $B'$ and $D'$.
A free vortex at $O$ provides a finite value of the fluid velocity on the apex of the wedge in the 
case of sideslip.

 With reference to the $\zeta$-plane, the complex potential must satisfy the following conditions: $f$ must be real on the arc 
$DAB$ and its real part must monotonically increase from $A$ to $B$ and from $A$ to $D$, and go to infinity at $C$ and $O$. 
\begin{figure}
\vspace{0.mm}
\centering
\includegraphics[width=0.90\textwidth]{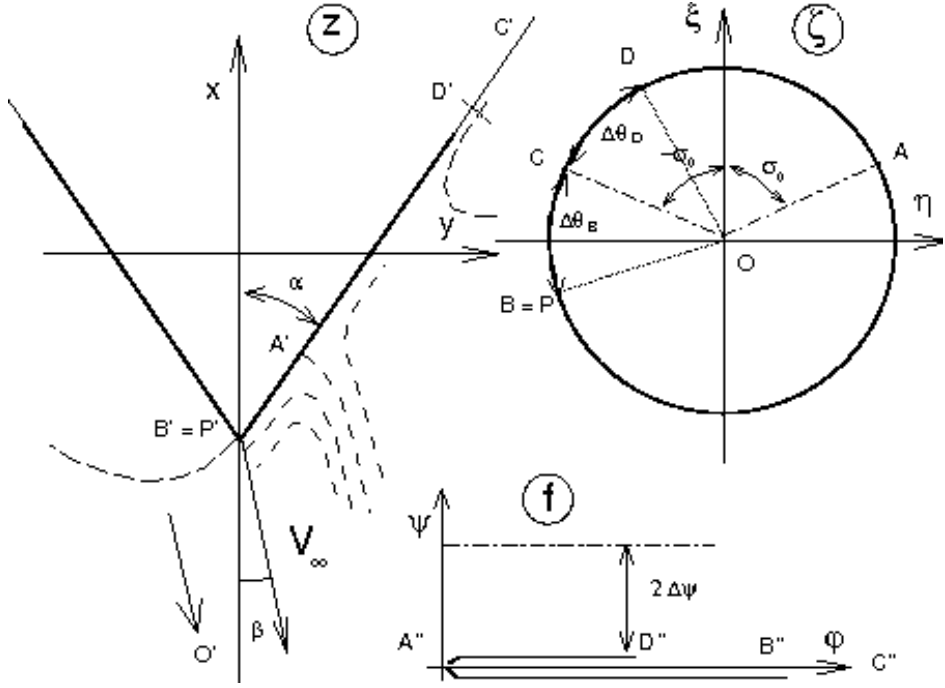}
  \caption{Conformal mapping (separated flow).}
\end{figure}
Points $A$, $B$, $C$ and $D$ will be transformed into $A''$, $B''$, $C''$ and $D''$ on the $f$-plane where  a cut is present on the 
horizontal axis $\varphi$ with the two branches corresponding to the two sides of the same iso-$\psi$ line which are divided 
by the stagnation point $A'$. Points $B$ and $D$, to be determined, represent the traces of the line along which $C_p=0$ on the wedge.
Finally, the angles $\Delta {\vartheta}_B$ and $\Delta {\vartheta}_D$ correspond to the arcs $BC$ and $CD$, respectively.     
Since the transformed free-surface lines on $\zeta$ are lines which connect $O$ with $B$ and $C$, the transformed domain is
the finite region bordered by the unit circle less the sector between $B$, $O$ and $D$.

The complex potential that satisfies all the required conditions will be  obtained by applying the Schwarz-Christoffel method.
 Let us then assume an expression for $f$ which is the sum of five terms
\be
\ds
 f \ = {\lambda_0} +  \ {\lambda_1} \ \frac{i}{2} \ds (\zeta - \frac{1}{\zeta})   +
 \  \frac{\Delta \psi} {\pi} \ln (1-\frac{1}{2} \ i (q \zeta - \frac{1}{q \zeta}) )   \\
+  \ds \chi \ds \frac{ \ds q \zeta + {1}/{q \zeta}}{ 1- \frac{1}{2} \ds  i (q \zeta - {1}/{q \zeta}) }  
- \ i \ \Gamma \ \ln  \zeta,  
\label{Chris}
\ee
\no where 
$    
{\sigma}_0 = \ds {\pi}/{2} \ (1+ \ds {\beta}/{\alpha}) \nonumber
$   
and 
$    
\ds q = \exp(i ({\sigma}_0- \ds {\pi}/{2}) ). \nonumber
$  
 
The five  terms in (\ref{Chris}) are $(a)$ a constant $\lambda_0$  which is associated to 
the presence of a free surface; $b)$ a doublet placed at $O$ of intensity $\lambda_1$ which corresponds to the translational
flow in the physical plane; $(c)$ a sink  at $C$ of intensity $2 \Delta \psi = {\Delta \psi}_B + {\Delta \psi}_D$, 
that depends on the angle $\alpha$ and on the sideslip $\beta$; $(d)$ a distorted doublet, also at $C$, that depends on $\beta$;
 $(e)$ a free vortex of intensity $\Gamma$ on the border at  $O$.  $\chi$ and $\Gamma$ are zero in the symmetric case.
The sum of the flow rates of the two jets corresponds to  2 $\Delta \psi$. When the flow is not symmetric, the two flow 
rates ${\Delta \psi}_B$ and ${\Delta \psi}_D$ are different and, in terms of the complex potential, this
partition is associated with the doublet $\chi$ in (\ref{Chris}). The presence of $\Gamma$ at infinity in $O$ does not yield
vorticity in the field. 

The properties of the complex velocity $\ds w = \ds $d$f/$d$ \zeta \ \ds $d$ \zeta/$d$z = u -$i$ v$
on the physical plane $z$ need to be considered in order to complete the formulation of the problem and to recover the shape
of the free surface once the problem in the transformed plane has been solved. In this respect 
it is useful to express the complex velocity through the so-called Levi-Civita auxiliary function  $\omega(\zeta)$
(see Cisotti 1921) such that $w(\zeta) = \exp[- $i$ \omega(\zeta)]$. In particular $w$ must be zero at $A$ and its
 modulus must increase monotonically from $A$ to $B$ and from $A$ to $D$, taking a  unit value at $O$, and its real and imaginary 
parts must satisfy the condition of null normal velocity  on the circle on $\zeta$.

The auxiliary function $\omega(\zeta)$ which satisfies the kinematic condition at the wall of the wedge is obtained by 
 the Schwarz-Christoffel formula
\bea
\omega(\zeta)=\beta -\alpha + \frac{2 i \alpha}{\pi} \log ( \frac{\zeta -j}{1-j \zeta}); \ \ j=\exp(i \sigma_0 ).   
\label{omega}
\eea
All the points on the physical plane are then obtained through the quadrature formula
\bea
\ds z(\zeta) = z(\zeta_0) + \  {\int_{\zeta_0} }^{\zeta}  \  e^{i \omega(\zeta)}  \ \frac{df(\zeta)} {d\zeta} \ d \zeta
\label{zita}
\eea
that gives the anti-transformation of the $\zeta$-plane onto the physical plane.

Returning to the evaluation of the five unknown constants, reference must be made to conditions which are meaningful from
the physical point of view.
First, mass continuity, momentum balance and energy conservation must be satisfied. Moreover  a condition
 (to be discussed later) must be imposed on the flow characteristics around the wedge apex when $\beta \ne 0$. 
 
Volume conservation takes into account the contributions of the free-surface displacement from the unperturbed free surface, 
as the wedge penetrates into the liquid, and the flow rates of the jets.
The free surface is like a material surface along which the pressure coefficient $C_p$ is zero, that is not perturbed
 at infinity and that ends at the contact iso-$\psi$ lines of the lateral jets.
Let $A_i$ be the immersed  float area.  The pertinent condition is 
\bea
R_1 \equiv {\int_B}^{\infty} \bar{x}(s) \ \frac{d\bar{y}(s)}{ds} \ ds \ + 
           {\int_D}^{\infty} \bar{x}(s) \ \frac{d\bar{y}(s)}{ds} \ ds  +
           \frac{{\Delta \psi}_B +{\Delta \psi}_D}{2}  - A_i = 0.
\label{R1}
\eea
The Bernoulli theorem, following the relations (\ref{similar}) and (\ref{similarV}), states that
$
\ds {\partial \Phi}/{\partial t} = (\varphi -{\bf r} \cdot {\bf v}) {V_{\infty}}^2
$. 
Then the expression for the pressure coefficient becomes
$
 {C_p} = 1- {\bf v \cdot \bf v } - \ 2 (\varphi - {\bf r} \cdot {\bf v}) \nonumber                                
$,
where $\bf r$ is the distance of a point ($x, y$) of the flow field  from the apex $P'$.    
Therefore the condition to be imposed at all points of the free surface is
\bea 
  1- {\bf v \cdot \bf v } - \ 2 (\varphi - \bar{\bf r} \cdot {\bf v}) =0,
\label{free_surf}                            
\eea 
where $\bar {\bf r}$ is the distance of a point of the free surface  ($\bar{x}, \bar{y}$) from $P'$.    

 Two further scalar conditions come from  the momentum balance and state that the integral of the pressure distribution on the body 
surface $S_B$ must be equal to the time-derivative of the momentum, including the effects of the lateral jets
\bea
(R_2, R_3) \equiv \int_{S_B} (p-p_{\infty})  {\bf n} \ dS + \frac{d}{dt} \int_S \rho \ \varphi \ {\bf n} \ dS  
+ \Delta {\bf q}_{B'} + \Delta {\bf q}_{D'}   =0.
\label{R23}
\eea
where $\Delta {\bf q}_{B'}$ and $\Delta {\bf q}_{D'}$ are the contributions of the lateral jets.
Since the structure of a jet is modeled as a sink through which mass, momentum and energy
disappear then, in particular, the lost momentum $\Delta{\bf q}$ and energy  $\Delta {E}$ can be
evaluated by means of the developed jet approximation
\bea
\Delta {\bf q} = \int_{\Delta \psi} ({\bf v} -{\bf r} ) \ d \psi \ \approx \ \Delta \psi \ (V -r)  \ {\bf t}
\nonumber
\eea
\bea
\Delta {E} = \frac{1}{2} \int_{\Delta \psi} ({\bf v} -{\bf r} ) \cdot ({\bf v} -{\bf r} ) \ d \psi \ \approx  \
{\small \frac{1}{2}} \Delta \psi \ (V- r)^2
\nonumber
\eea
where {\bf t} is the unit vector parallel to the wedge side.
At small deadrise angle $\delta$ the jet approximation is satisfied in the limit $\delta \rightarrow 0$
while at $\delta \rightarrow 90 ^{\circ}$ both the exact and the approximate expressions  
$\Delta{\bf q} \rightarrow 0$, $\Delta E \rightarrow 0$, since $\Delta{\psi} \rightarrow 0$.
At intermediate deadrise angle the jet approximation might involve greater errors but the calculation
performed and presented here give results which are in excellent agreement with the existing data of
\cite{Zhao93}.

With reference then to (\ref{R23}) one has 

\bea 
\Delta {\bf q}_{B'} = \ds \int_{\Delta \psi_{B'}} ({\bf v} -{\bf r})  \ d \psi \approx \Delta {\psi}_{B'} (V_{B'}-r_{B'}) {\bf t}_{B'}
\nonumber
\eea
 and 
\bea
\Delta {\bf q}_{D'} = \ds \int_{\Delta \psi_{D'}} ({\bf v} -{\bf r})  \ d \psi \approx \Delta {\psi}_{D'} (V_{D'}-r_{D'}) {\bf t}_{D'},
\nonumber
\eea
where, as before, ${\bf t}_{B'}$ and  ${\bf t}_{D'}$ are the local tangent unit vectors, and 
$r_{B'}$ and $r_{D'}$ are the distances from the apex of the wedge to the points $B'$ and $D'$, respectively.
  
Note that the derivative of the momentum for a constant entry velocity is due to the variation of the apparent additional mass.
Note also that in the proper dimensionless similarity variables $r_{B'}$ and $r_{D'}$ are also the velocities of the points $B'$ and $D'$.

The lateral surface of the wedge depends on $\Delta \vartheta_B$ and $\Delta \vartheta_D$ through 
\bea
\ds S_B(\Delta \vartheta_B, \Delta \vartheta_D) = 
{\ds \huge \int}_{\sigma_0 + \Delta \vartheta_B}^{2 \pi -\sigma_0 - \Delta \vartheta_D} 
\vert \ds \frac{d f }{d \zeta} \vert  \ \ds \frac{1}{\vert w \vert }  d \vartheta. \nonumber
\eea
and $S$ is where $S_B$ meets the free surface.

 The energy conservation takes into account the presence of the free surface and of the jets and equates  the work of the hydrodynamic
 force on the body to the time derivative of the kinetic energy: 
\bea
R_4 \equiv  \int_{S_B} (p-p_{\infty}) \ {\bf V_{\infty}} \cdot {\bf n} \ dS - 
\frac{1}{2} \frac{d}{dt}  \int_S \rho \ \varphi \ \nabla \varphi \cdot {\bf n} \ dS + 
\Delta E_{B'} + \Delta E_{D'} = 0,
\label{R4}
\eea
where, as before,
\bea
\Delta E_{B'} = {\ds \frac{1}{2}} \int_{\Delta \psi_{B'}} ({\bf v} -{\bf r}) \cdot ({\bf v} -{\bf r})  \ d \psi \approx \frac{1}{2} \Delta {\psi}_{B'} (V_{B'}-r_{B'})^2
\nonumber
\eea
 and 
\bea
\Delta E_{D'} = \ds \frac{1}{2} \int_{\Delta \psi_{D'}} ({\bf v} -{\bf r}) \cdot ({\bf v} -{\bf r}) \ d \psi \approx \ds \frac{1}{2} \Delta {\psi}_{D' }(V_{D'} -r_{D'})^2 
\nonumber
\eea
are the kinetic energy terms which are lost through the lateral jets.

It is worth remarking that since the expressions for $\Delta \bf q$ and $\Delta E$ were obtained
under the approximation of a developed jet, they do not allow a fully detailed
description of the flow structure near the walls.

 Whereas for $\beta =0$ the stagnation point $A'$ falls on the apex, two possible choices can instead be made for $\beta \ne 0$:
either point $A'$ is still coincident with the apex (figure 2) and no separation occurs, or the free surface separates downstream 
from the apex, and $A'$ moves along the upstream wet side of the wedge (figure 3). In both cases one has the further 
condition  $\ds \frac {df} {d \zeta} = 0$ in $\zeta_A$, i.e.
\bea
R_5 \equiv \Gamma - \lambda_1 \cos \sigma_0 +\frac{\Delta \psi /\pi \cos(\sigma_0 + \varepsilon)+
2 \ \chi }{1+\sin(\sigma_0 + \varepsilon)} = 0,
\label{R5}
\eea
\no where $\varepsilon=\sigma_0 - \pi/2$.
In the expressions for $R_1$ to $R_5$ above a few auxiliary unknowns appear, namely the angles 
$\Delta {\vartheta}_B$, $\Delta {\vartheta}_D$ and the jumps of the streamfunction $\Delta \psi_{B'}$ and $\Delta \psi_{D'}$.

Let us discuss the attached case first.
The conditions  
\bea
R_6 \ \equiv {C_p}_{B'} =0 ,  \ \  R_7 \equiv {C_p}_{D'} =0  
\label{R67}
\eea
give the values of $\Delta {\vartheta}_B$ and $\Delta {\vartheta}_D$ and the total flow rate of the two jets $2 \Delta \psi$ 
is divided in such a way that
\bea
R_8 \equiv 2 \Delta \psi - ( \Delta \psi_{B'} + \Delta \psi_{D'} ) = 0.
\label{R8}
\eea

As has been seen, in the case of attached flow the volume flow rates of the lateral jets correspond to two jumps of the streamfunction.
The Bernoulli equation for each contact iso-$\psi$ line which comes from infinity gives
\bea
R_9 \equiv \frac{\Delta \psi_{B'}}{\Delta \psi_{D'}} - \frac{{V_{B'}}^2 - 1}{ {V_{D'}}^2 - 1} = 0,
\label{R9}
\eea
where $V_{B'}$ and $V_{D'}$ are evaluated as $\vert w({\zeta}_B) \vert$ and $\vert w({\zeta}_D) \vert$ respectively, 
since the potential time derivatives are recognized to be 
    $\ds ({\partial \varphi}/{\partial t}) {\vert}_{B'} = \ $const$ \ \Delta \psi_{B'} $
and $\ds ({\partial \varphi}/{\partial t}) {\vert}_{D'} = \ $const$ \ \Delta \psi_{D'} $.

When we consider the case of a separated flow, the problem reduces to that of a flat plate entering the water at an angle $2 \pi - (\alpha +\beta)$.
One of the two streamfunction jumps is zero and instead of (\ref{R9}), the condition that the apex of the wedge  
is a point of separation is applied:
 \bea
\ds R_9 \equiv \frac{d f(\zeta_P)}{d \zeta} = 0. 
\label{R10}
\eea

We finally turn our attention to the shape of the free surface in the physical plane.
The kinematic condition on that surface is
\vspace{-5 mm}
\bea
\frac {d \bf r} {ds} \ = \ \frac {\bf r - \bf v} {\sqrt{(x-u)^2+(y-v)^2} },
\label{rs}
\eea
where (\ref{free_surf}) must be satisfied also.
The vector equation (\ref{rs}) expresses the fact that the free surface is always made up of the same
particles. Equation (\ref{rs}) when combined with (\ref{free_surf}) 
confirms that the arc distance between two particles on the surface is constant (see, for instance, 
Birkhoff $\&$ Zarantonello 1957, Mackie 1962).

Introducing the complex notation we obtain, in the transformed variables, the quadrature equations
from (\ref{rs}) and (\ref{free_surf}):
\bea
z(\zeta)  = z_i + {\int_{\zeta_i}}^{\zeta} 
\frac {z(\zeta') - w^*(\zeta')} {\sqrt{(x(\zeta')-u(\zeta'))^2+(y(\zeta')-v(\zeta'))^2} }
 {\mid z_{\zeta}(\zeta') \mid }  d{\mid \zeta' \mid},  \ (i = B', D')    
\label{zint}
\eea
where the integrals  are restricted to the paths on $\zeta$ that correspond to  $C_p = 0$. The most difficult part
 of the differential problem,  i.e. the determination of the shape of the free surface, is thus formally reduced to 
the integration of  expressions (\ref{zint})  along the paths along which  $C_p = 0$.

\section {\bf  The algorithm }

Expressions (\ref{R1})-(\ref{R9}) (or (\ref{R1})-(\ref{R8}), \ref{R10})) represent an algebraic system with nine unknown parameters.
Among the various solution methods, a speedy and easy way was adopted which corresponds to solving the problem in the form of a residual 
function
 
\hspace{2.5cm}$
\ds R(  \lambda_0, \ \lambda_1, \ \Delta \psi, \ \chi, \ \Gamma,  \ \Delta \psi_B, \ \Delta \psi_D,  \ \Delta \vartheta_B, \ \Delta \vartheta_D )= \sum_{i=1}^9  R_i^2 
$

\no which is made zero through an optimization procedure.
In particular the optimization method by Davidon, Fletcher and Powell was adopted which is a numerical 
code in the widely known $NAG$ library (Numerical Algorithms Group 1991 ). 

In detail, the steps of  the algorithm  to obtain the solution in the physical plane as follows. 

\begin{enumerate}[(iii)]
\renewcommand{\theenumi}{\roman{enumi}}
\item  The analytical form of the complex potential (\ref{Chris}) that satisfies all the required
boundary conditions is assumed.  

\item As a first guess values of all the  nine parameters $\lambda_0$, $\lambda_1$, $\Delta \psi$, $\chi$, $\Gamma$, $\Delta \psi_B$, 
$\Delta \psi_D$,  $\Delta \vartheta_B$, $\Delta \vartheta_D$ are chosen.
Therefore the complex potential velocity is completely defined in the  $\zeta$-plane.

\item All the points on the physical plane are calculated via (\ref{zita}), so that a first approximation to 
the solution which describes the flow around the wedge is evaluated.

\item If the solution so obtained satisfies the algebraic set and  makes function $R$ go to zero,
the procedure is terminated and the physical solution has been found. Otherwise, the optimization 
process will update the nine parameters in such a way that $R$ is minimized 
and the procedure will be repeated from step {(iii)}.

\item Once the parameters which comprise $\ds \sum_{i=1}^9  R_i^2 = 0$ are evaluated, next step is to calculate
the  shape of the free surface through integration of (\ref{zint}). To this end, in this work a routine 
fourth-order Runge-Kutta method was adopted.

\end{enumerate}

\begin{figure}
\vspace{0.mm}
\centering
\includegraphics[width=0.90\textwidth]{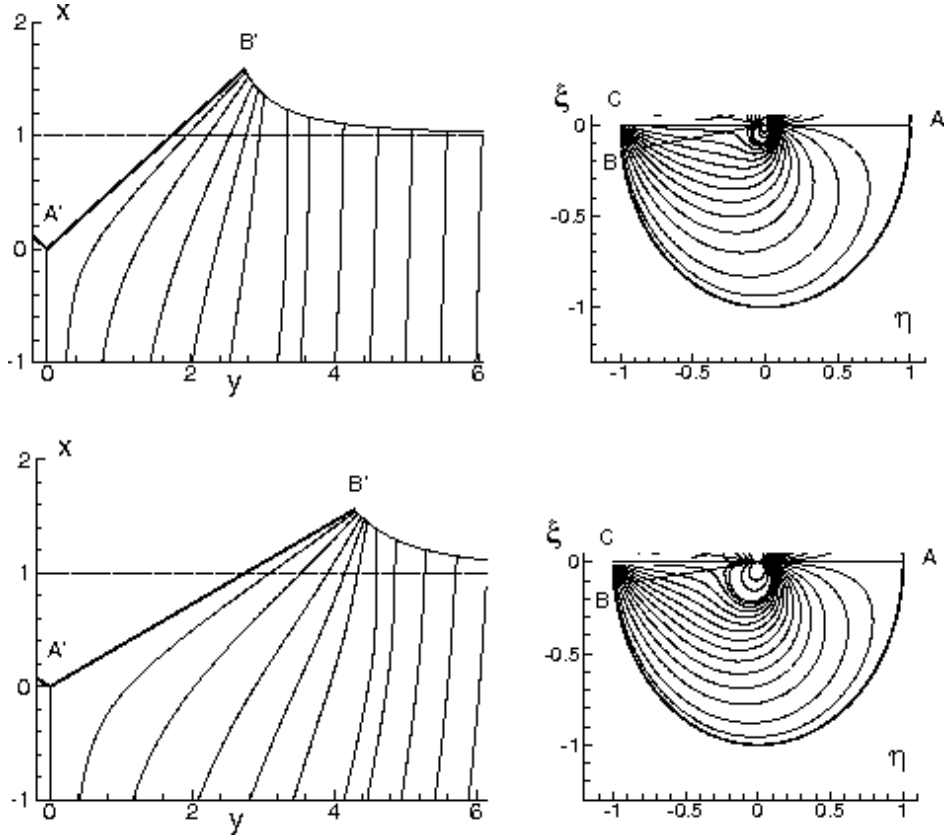}
  \caption{ Symmetrical flow field, $\alpha = 60 ^{\circ}$ (a), $\alpha = 70 ^{\circ}$ (b).}
\end{figure}

\section {\bf  Results }

Some results will now be presented and discussed. First, some solutions
obtained in this paper were substituted, for comparison, into the singular integral equation obtained in \cite{Dobrovolskaya69}
for the symmetric case and it was solved by a finite difference method.
In particular, for $\beta=0$, we considered the cases where  $\alpha= \ 60^{\circ}$ and  $70^{\circ}$ which were thought to 
represent good tests for the solution procedure. 
In fact these intermediate values of $\alpha$ correspond to situations which are far from the limits where the existing 
approximate solutions are accurate enough. Therefore 
our solutions were substituted into Dobrovol'skaya's integral equation 
at a number of points of the liquid domain and of the wetted surface and the results with negligible 
differences corresponded to the data reported in \cite{Dobrovolskaya69} as corrected by \cite{Zhao93}.

After proving the reliability of the method, the situations considered in \cite{Zhao93}
which were solved there by a nonlinear boundary elements method were dealt with by the present approach.
Again, an  excellent agreement was found between the two procedures, apart from negligible numerical errors. 
\begin{figure}
\vspace{0.mm}
\centering
\includegraphics[width=0.90\textwidth]{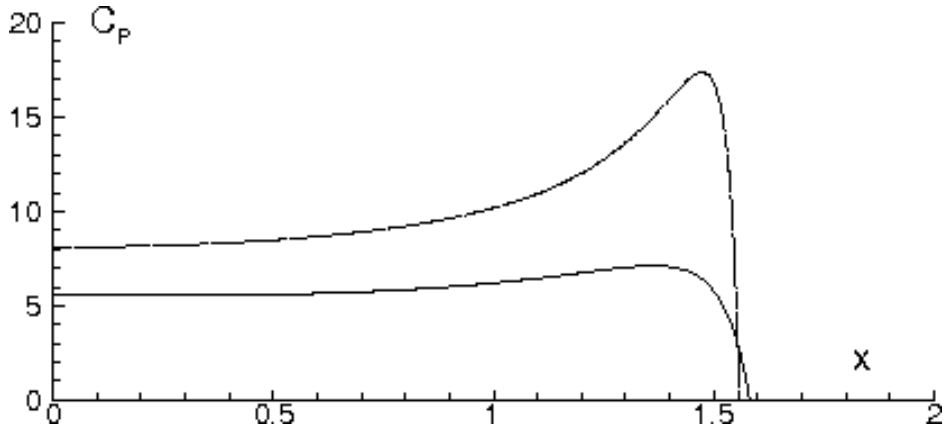}
  \caption{Symmetrical pressure distribution, $\alpha = 60 ^{\circ}$ (continuous line),\protect\\ $\alpha = 70 ^{\circ}$
 (dashed line). }
\end{figure}
The results for the cases cited above are given in figure 4 where some iso-$\psi$ lines are sketched both in the physical 
and in the transformed  plane.
The calculated $C_p$ vs. $x$ along the walls is shown in figure 5. The calculations proved that the present analysis
provides excellent solutions in a very simple and fast way over the entire range of $\alpha$, the case $\alpha =0$ and $\alpha =90^{\circ}$
being excluded.  At smaller deadrise angles than those in figures 4 and 5, a comparison was carried between our $C_p$ results and those shown
in \cite{Zhao93} for $\alpha = 86 ^{\circ}$. In this case we note a maximum deviation of about $10 \%$ which occurs at the wall. 
\begin{table} 
\hspace{3.0cm} ${C_p}_{max}$     \hspace{6.0cm}  $(x_{max}-1)$
\vspace{-0.6cm}
  \begin{center}
  \begin{tabular}{lccccccccc} 
      $\alpha (deg.)$      &Simil.    & Asymp.   & BE    & Present results \hspace{1.00cm} &Simil.  & Asymp.  & BE    & Present results\\[1pt] 
      50                   & 3.266    & 3.50     & 3.26  & 3.5260  \hspace{1.00cm} & 0.2866 & 0.5708  & 0.245 & 0.2687\\ 
      60                   & 6.927    & 7.40     & 6.94  & 7.1127  \hspace{1.00cm} & 0.4243 & 0.5708  & 0.400 & 0.3655\\ 
      65                   & 10.691   & 11.35    & 10.9  & 10.622  \hspace{1.00cm} & 0.4709 & 0.5708  & 0.443 & 0.4257\\ 
      70                   & 17.774   & 18.63    & 18.2  & 17.370  \hspace{1.00cm} & 0.5087 & 0.5708  & 0.488 & 0.4728\\ 
      75                   & 33.271   & 34.37    & 32.8  & 32.654  \hspace{1.00cm} & 0.5361 & 0.5708  & 0.533 & 0.5158\\
      80                   & 77.847   & 79.36    & 80.2  & 78.805  \hspace{1.00cm} & 0.5556 & 0.5708  & 0.555 & 0.5619\\
      82.5                 & 140.587  & 142.36   & 148.3 & 146.422 \hspace{1.00cm} & 0.5623 & 0.5708  & 0.558 & 0.5691\\
      86                   & 503.030  & 504.61   & 521.4 & 512.324 \hspace{1.00cm} & 0.5695 & 0.5708  & 0.571 & 0.5708\\ 
  \end{tabular}
  \end{center}
\vspace{0.5cm}
\hspace{4.5cm} $\ds F_x$
  \begin{center}
 \begin{tabular}{lccccccc} 
      $\alpha (deg.)$      &Simil.    & Asymp.   & BE    & Present results \hspace{1.0cm} & $E_{kin}$    & $E_{jets}/E_{kin}$\\[1pt] 
      50                   & 5.477    & 8.322    & 5.31   & 5.7154   \hspace{1.0cm} &  3.541     &  0.6140\\ 
      60                   & 14.139   & 18.747   & 13.9   & 16.168   \hspace{1.0cm} &  9.424     &  0.7154\\ 
      65                   & 23.657   & 29.765   & 23.7   & 25.829   \hspace{1.0cm} &  14.448    &  0.7877\\ 
      70                   & 42.485   & 50.639   & 43.0   & 44.020   \hspace{1.0cm} &  23.714    &  0.8562\\ 
      75                   & 85.522   & 96.879   & 85.5   & 83.841   \hspace{1.0cm} &  43.756    &  0.9161\\
      80                   & 213.98   & 231.973  & 220.8  & 202.125  \hspace{1.0cm} &  104.230   &  0.9622\\
      82.5                 & 399.816  & 423.735  & 417.9  & 401.343  \hspace{1.0cm} &  192.78    &  0.9788\\
      86                   & 1503.638 & 1540.506 & 1491.8 & 1487.321 \hspace{1.0cm} &  744.14    &  0.9987\\ 
\end{tabular}
\caption{Dimensionless slamming parameters vs. $\alpha$. Comparison of maximum pressure coefficient and its location
 $x_{max}$, and of vertical force $F_x$  as obtained in a similar solution,  an asymptotic analysis, a
 nonlinear boundary element method BE and the present procedure.  
$E_{kin}$, the kinetic energy of the bulk of the fluid, and $E_{jets}$, kinetic energy lost in the jets are also shown.
 Data for all but the present results are from (\cite{Zhao93})}.
\end{center} 
\end{table} 
\begin{table} 
  \begin{center} 
  \begin{tabular}{lcc} 
      $\beta (deg.)$      & Present results  \    & von K\'arm\'an results  \\[1pt] 
       4 & 5.415   & 4.692\\
       6 & 5.382   & 4.664\\
       8 & 5.336   & 4.620\\
      10 & 5.277   & 4.573\\
  \end{tabular}
  \caption{Values of the total force on the wet side of the wedge for a separated flow, $\alpha = 60^{\circ}$.}
  \end{center} 
\end{table} 
Table 1 presents a comparison of the results of the present procedure with those obtained by a similarity solution, an asymptotic
approach and the boundary element method. In addition the calculated values of the kinetic energy of the flow and of the jets
are shown. 
One can see the increasing importance of the momentum and energy associated with the jets as $\alpha$ increases.

On the specific point of the energy balance for the impact of a symmetric body, a general discussion appears in \cite{Molin96}.
In accordance with this last reference the energy going into the jets tends to be equal to the kinetic energy of the bulk of 
the fluid as the deadrise angle decreases.

When the case with a sideslip is considered one finds that the solution for the attached case is not always possible. In particular,
a limit value of $\beta$ exists for each $\alpha$, namely ${\beta}^*$, such that for $\beta > \ {\beta}^*$ the only possible solution 
corresponds to separated flow. In order to obtain ${\beta}^*$ from (\ref{R1})-(\ref{R9}) we calculated the Jacobian 
$\ds {\partial R_i}/{\partial q_j}$ corresponding to the solutions of the problem. Here $q_j$ represents
 the generic variable on which $R$ depends. Then ${\beta}^*$ is obtained when the determinant of $\ds {\partial R_i}/{\partial q_j}$
vanishes. As expected and with reference to figure 6 we note that ${\beta}^*$ increases with $\alpha$.
\begin{figure}
\vspace{0.mm}
\centering
\includegraphics[width=0.90\textwidth]{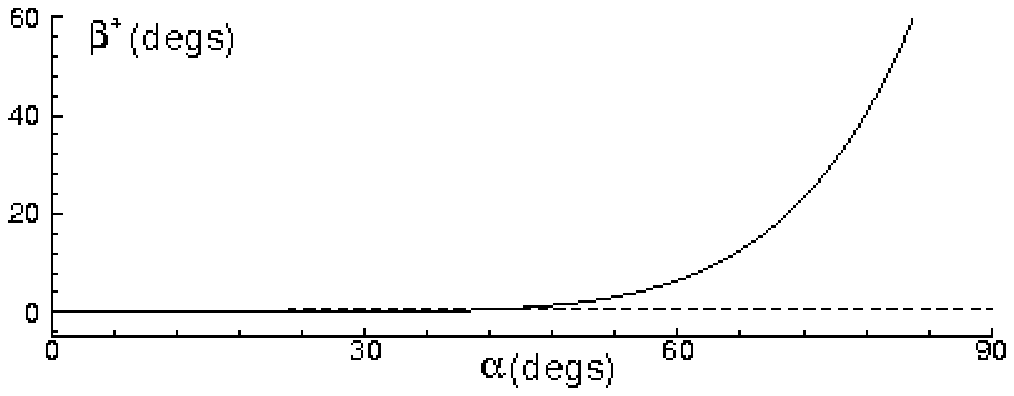}
  \caption{Values of the sideslip angle for the onset of separation ${\beta}^*$ vs. the wedge angle $\alpha$.
Above the line the flow separates from one side.}
\end{figure}

Figure 7 shows the solutions of the flow field in the case 
of sideslip $\beta = 4 ^{\circ}$, for $\alpha =60 ^{\circ}$ and $70 ^{\circ}$, when the flow is attached to both walls. 
The corresponding distributions of the pressure coefficient are given in figure 8 together with the results for the symmetric case. 
On the windward wall the free surface rises higher than on the leeward wall and on the leeward side the pressure coefficient reaches 
a maximum value which is close to but greater than that of the symmetric case.
\begin{figure}
\vspace{0.mm}
\centering
\includegraphics[width=0.90\textwidth]{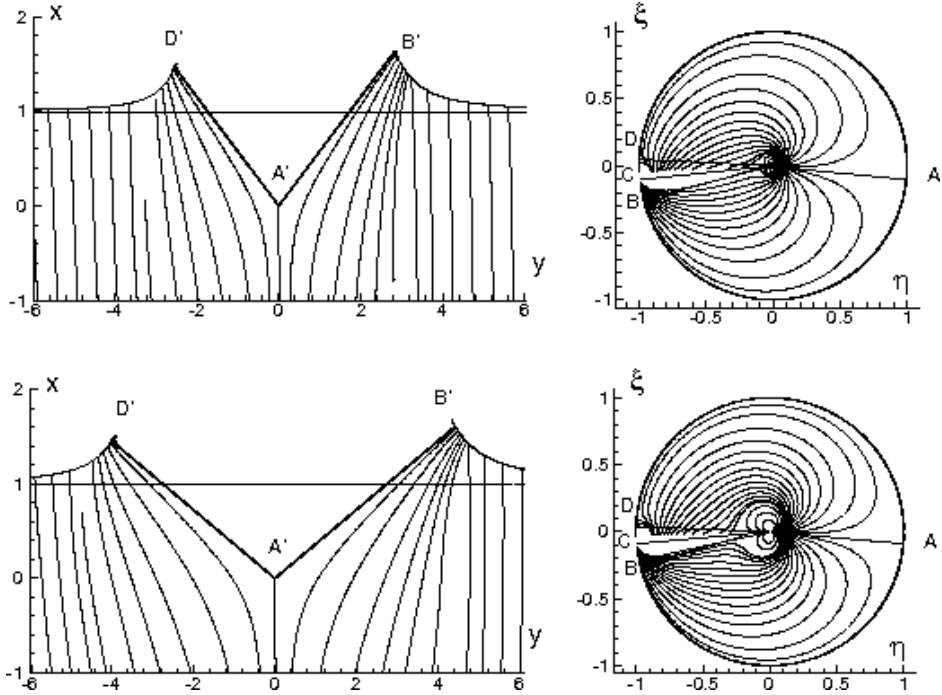}
\caption{Unsymmetrical attached flow field: iso-$\psi$ line distributions in the physical and in the transformed plane, 
$\alpha = 60 ^{\circ}$, $\beta = 4 ^{\circ}$  (a); $\alpha = 70 ^{\circ}$, $\beta = 4 ^{\circ}$ (b).}
\end{figure}
When the flow separates from the wedge then, as already mentioned, the problem reduces to that of a flat plate which enters the liquid 
at an angle $\beta$. Figures 9 and 10 show the flow field for $\beta = 4 ^{\circ}$ and $\alpha = 60 ^{\circ}$,
and the corresponding distribution of the pressure coefficient, respectively. In comparison with the symmetric case both the
 maximum value ${C_p}_{max}$ and the entire $C_p$ distribution show noticeable differences.
Furthermore the free surface on the downstream side and close to the wall falls below the unperturbed surface. 

When the inertia force is calculated from the displaced virtual mass, one obtains values which 
approximate within a $15 \%$ error the \cite{Karman29} results  which are evaluated with reference 
to the mass contained in a cylindrical volume having diameter equal to $\tan \alpha$.

Figure 10  shows  a comparison between the pressure coefficient distributions on the wall for 
$\alpha = 60 ^{\circ}$ and $\beta = 10 ^{\circ}$ as calculated by our method and by von K\'arm\'an's  approximation.
In table 2, for $\alpha = 60^{\circ}$ and at different $\beta$,
the total force on the wet side of the wedge is reported together with the corresponding von K\'arm\'an values.
Higher-order approximations of the flat-plate virtual mass, such as the one discussed in \cite{Meyerhoff70}, are not
considered  here.
\begin{figure}
\vspace{0.mm}
\centering
\includegraphics[width=0.90\textwidth]{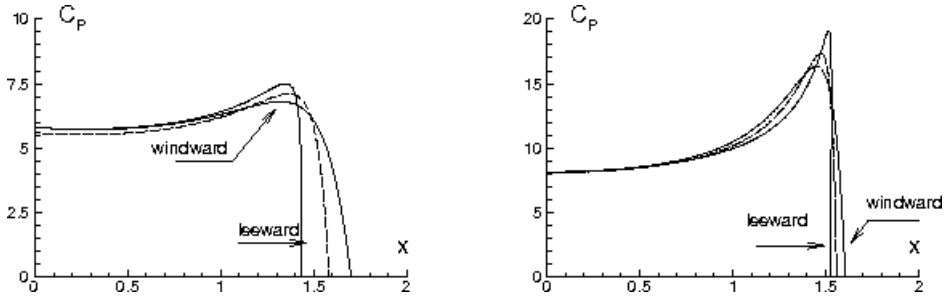}
  \caption{Pressure distribution for attached flow, $\alpha = 60 ^{\circ}$ (a), $\alpha = 70 ^{\circ}$ (b),\protect\\
 $\beta = 0$  (dashed lines), $\beta = 4 ^{\circ}$ (continuous lines).}
\end{figure}
\begin{figure}
\vspace{0.mm}
\centering
\includegraphics[width=0.90\textwidth]{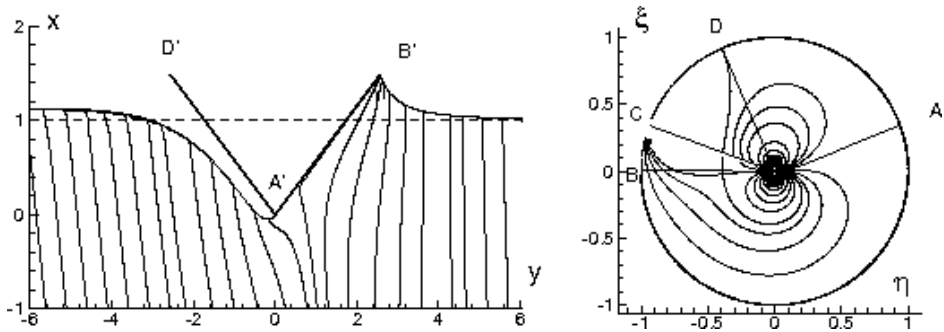}
  \caption{Separated flow field. $\alpha = 60 ^{\circ}$ and $\beta = 10 ^{\circ}$: 
iso-$\psi$ line distributions in the physical and in the transformed plane.}
\end{figure}
\begin{figure}
\vspace{0.mm}
\centering
\includegraphics[width=0.90\textwidth]{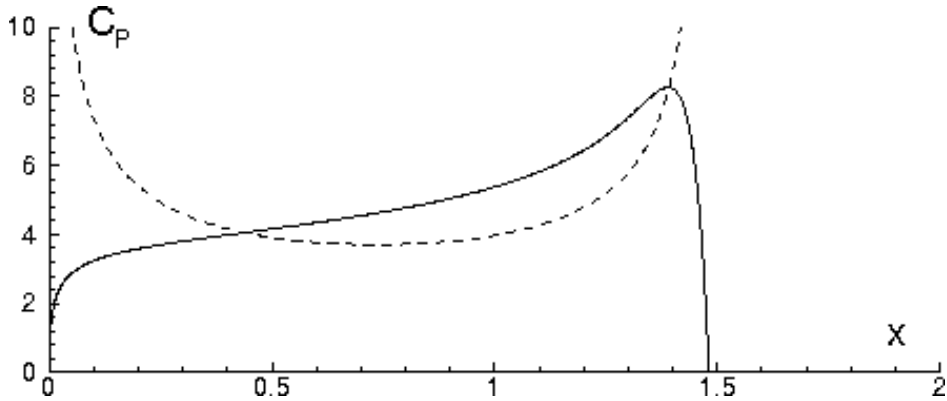}
  \caption{Pressure distribution for the separated  flow, $\alpha = 60 ^{\circ}$, 
 $\beta = 10 ^{\circ}$:  von K\'arm\'an's theory (dashed line), present method (continuous line)}
\end{figure}
\section {\bf  Conclusion}
We conclude by summarizing the characteristics of the present approach. 
The Laplace equation is supposed to hold, after the introduction 
of similar variables gives a steady expression for the flow field. 
Then the physical aspects are modelled by a suitable choice of 
singularities of the potential function in a translational potential.
 Following a conformal transformation, the main characteristics of 
the flow in the transformed plane are found by solving a system of algebraic equations for the singularities by an optimization
procedure and the shape of the free surface is formally obtained by quadrature, although an easy way to practically 
compute it is based on a Runge-Kutta method. 
An important aspect of the model is the fact that the law of mass continuity, momentum balance and energy conservation are enforced.
 The solutions were tested against existing data with excellent results. 
The method can be applied to the entire range of wedge angles, except for $\alpha = 0$ and $\alpha = 90 ^{\circ}$, and takes
easily into account a possible sideslip with and without separation. 

\vspace{0.500cm}

This work was partially supported by  
the Italian Ministry for 
the Universities and Scientific and Technological Research.

\end{document}